\documentclass[a4paper,12pt]{article}
\usepackage{graphics,psfrag}
\usepackage{amsmath,amsthm,amssymb,epsfig,euscript,array,cite,cancel,color}
\usepackage{cases,empheq}
\usepackage{verbatim}
\usepackage{mciteplus}

\if{}
\usepackage[paper=a4paper,dvips,top=2.54cm,left=2.74cm,right=2.34cm,
    foot=1cm,bottom=2.54cm]{geometry}
\fi

\theoremstyle{definition}

\theoremstyle{remark}

\newcounter{multieqs}

\newcommand{\be}{\begin{equation}}
\newcommand{\ee}{\end{equation}}
\newcommand{\eq}[1]{(\ref{#1})}
\newcommand{\bit}{\begin{itemize}}  \newcommand{\eit}{\end{itemize}}

\newcommand{\bm}[1]{\mbox{\boldmath $#1$}}
\newcommand{\rf}[1]{(\ref{#1})}

\def\bd{\begin{document}}
\def\ed{\end{document}}
\def\nn{\nonumber}
\def\bea{\begin{eqnarray}}
\def\eea{\end{eqnarray}}
\let\bm=\bibitem

\def\la{\langle}
\def\ra{\rangle}

\def\npb#1#2#3{Nucl. Phys. {\bf{B#1}} #3 (#2)}
\def\plb#1#2#3{Phys. Lett. {\bf{#1B}} #3 (#2)}
\def\prl#1#2#3{Phys. Rev. Lett. {\bf{#1}} #3 (#2)}
\def\prd#1#2#3{Phys. Rev. {D \bf{#1}} #3 (#2)}
\def\cmp#1#2#3{Comm. Math. Phys. {\bf{#1}} #3 (#2)}
\def\cqg#1#2#3{Class. Quantum Grav. {\bf{#1}} #3 (#2)}
\def\nppsa#1#2#3{Nucl. Phys. B (Proc. Suppl.) {\bf{#1A}}#3 (#2)}
\def\ap#1#2#3{Ann. of Phys. {\bf{#1}} #3 (#2)}
\def\ijmp#1#2#3{Int. J. Mod. Phys. {\bf{A#1}} #3 (#2)}
\def\rmp#1#2#3{Rev. Mod. Phys. {\bf{#1}} #3 (#2)}
\def\mpla#1#2#3{Mod. Phys. Lett. {\bf A#1} #3 (#2)}
\def\jhep#1#2#3{J. High Energy Phys. {\bf #1} #3 (#2)}
\def\atmp#1#2#3{Adv. Theor. Math. Phys. {\bf #1} #3 (#2)}

\def\N{{\cal N}}
\def\sst{\scriptscriptstyle}
\def\thetabar{\bar\theta}
\def\Tr{{\rm Tr}}
\def\one{\mbox{1 \kern-.59em {\rm l}}}

\def\a{\alpha}      \def\da{{\dot\alpha}}  \def\dA{{\dot A}}
\def\b{\beta}       \def\db{{\dot\beta}}
\def\g{\gamma}  \def\G{\Gamma}  \def\dc{{\dot\gamma}}
\def\d{\delta}  \def\D{\Delta}  \def\ddt{\dot\delta}
\def\e{\epsilon}        \def\ve{\varepsilon}
\def\f{\phi}    \def\F{\Phi}    \def\vvf{\f}
\def\h{\eta}
\def\k{\kappa}
\def\l{\lambda} \def\L{\Lambda}
\def\m{\mu} \def\n{\nu}
\def\o{\omega}
\def\p{\pi} \def\P{\Pi}
\def\r{\rho}
\def\s{\sigma}  \def\S{\Sigma}
\def\t{\tau}
\def\th{\theta} \def\Th{\Theta} \def\vth{\vartheta}
\def\X{\Xeta}
\def\z{\zeta}

\def\na{\nabla}

\def\cA{{\cal A}} \def\cB{{\cal B}} \def\cC{{\cal C}}
\def\cD{{\cal D}} \def\cE{{\cal E}} \def\cF{{\cal F}}
\def\cG{{\cal G}} \def\cH{{\cal H}} \def\cI{{\cal I}}
\def\cJ{{\cal J}} \def\cK{{\cal K}} \def\cL{{\cal L}}
\def\cM{{\cal M}} \def\cN{{\cal N}} \def\cO{{\cal O}}
\def\cP{{\cal P}} \def\cQ{{\cal Q}} \def\cR{{\cal R}}
\def\cS{{\cal S}} \def\cT{{\cal T}} \def\cU{{\cal U}}
\def\cV{{\cal V}} \def\cW{{\cal W}} \def\cX{{\cal X}}
\def\cY{{\cal Y}} \def\cZ{{\cal Z}}

\def\ua{\underline{\alpha}}
\def\uc{\underline{\phantom{\alpha}}\!\!\!\gamma}
\def\um{\underline{\mu}}
\def\ud{\underline\delta}
\def\ue{\underline\epsilon}
\def\una{\underline a}\def\unA{\underline A}
\def\unb{\underline b}\def\unB{\underline B}
\def\unc{\underline c}\def\unC{\underline C}
\def\und{\underline d}\def\unD{\underline D}
\def\une{\underline e}\def\unE{\underline E}
\def\unf{\underline{\phantom{e}}\!\!\!\! f}\def\unF{\underline F}
\def\unm{\underline m}\def\unM{\underline M}
\def\unn{\underline n}\def\unN{\underline N}
\def\unp{\underline{\phantom{a}}\!\!\! p}\def\unP{\underline P}
\def\unq{\underline{\phantom{a}}\!\!\! q}
\def\unQ{\underline{\phantom{A}}\!\!\!\! Q}
\def\unH{\underline{H}}

\def\As {{A \hspace{-6.4pt} \slash}\;}
\def\bs {{b \hspace{-6.4pt} \slash}\;}
\def\Ds {{D \hspace{-6.4pt} \slash}\;}
\def\Gts {{\Gt \hspace{-6.4pt} \slash}\;}
\def\ds {{\del \hspace{-6.4pt} \slash}\;}
\def\ss {{\s \hspace{-6.4pt} \slash}\;}
\def\ks {{ k \hspace{-6.4pt} \slash}\;}
\def\ps {{p \hspace{-6.4pt} \slash}\;}
\def\xs {{x \hspace{-6.4pt} \slash}\;}
\def\pas {{{p_1} \hspace{-6.4pt} \slash}\;}
\def\pbs {{{p_2} \hspace{-6.4pt} \slash}\;}
\def\cFs {{{\cal F} \hspace{-6.4pt} \slash}\;}

\def\Ah{{\hat{A}}}
\def\Dh{{\hat{D}}}
\def\Gh{{\hat{G}}}
\def\Fh{{\hat{F}}}
\def\Ih{{\hat{I}}}
\def\Jh{{\hat{J}}}
\def\Kh{{\hat{K}}}
\def\Lh{{\hat{L}}}
\def\Ph{{\hat{P}}}
\def\Rh{{\hat{R}}}
\def\Vh{{\hat{V}}}
\def\Xh{{\hat{X}}}

\def\ah{{\hat{a}}}
\def\bh{{\hat{b}}}
\def\ch{{\hat{c}}}
\def\gh{{\hat{g}}}
\def\dh{{\hat{d}}}
\def\hh{{\hat{h}}}
\def\uh{{\hat{u}}}
\def\vh{{\hat{v}}}
\def\xh{{\hat{x}}}
\def\yh{{\hat{y}}}
\def\zh{{\hat{z}}}
\def\ph{{\hat{p}}}
\def\thh{{\hat{t}}}
\def\xih{\hat{\xi}}
\def\Psih{\hat{\Psi}}
\def\mh{{\hat{m}}}
\def\nh{{\hat{n}}}
\def\ih{{\hat{i}}}
\def\jh{{\hat{j}}}
\def\kh{{\hat{k}}}
\def\aah{{\hat{\alpha}}}
\def\bbh{{\hat{\beta}}}
\def\ggh{{\hat{\gamma}}}
\def\llh{{\hat{\ell}}}
\def\ph{{\hat{p}}}

\def\psit{\tilde{\psi}}
\def\Psit{\tilde{\Psi}}
\def\Psibt{\tilde{\bar{Psi}}}

\def\st{\tilde{\sigma}}

\def\delt{\tilde{\delta}}
\def\Phit{\tilde{\Phi}}
\def\Phitb{\overline{\tilde{Phi}}}
\def\tht{\tilde{\th}}
\def\lt{\tilde{\l}}
\def\chit{\tilde{\chi}}
\def\phit{\tilde{\phi}}

\def\At{\tilde{A}}
\def\Bt{\tilde{B}}
\def\Ct{\tilde{C}}
\def\Dt{\tilde{D}}
\def\Et{\tilde{E}}
\def\Ft{\tilde{F}}
\def\Gt{\tilde{G}}
\def\Ht{\tilde{H}}
\def\It{\tilde{I}}
\def\Jt{\tilde{J}}
\def\Qt{\tilde{Q}}
\def\Rt{\tilde{R}}
\def\Mt{\tilde{M }}
\def\Nt{\tilde{N}}
\def\St{\tilde{S}}
\def\Vt{\tilde{V}}
\def\Xt{\tilde{X}}
\def\at{\tilde{a}}
\def\ct{\tilde{c}}
\def\dt{\tilde{d}}
\def\htt{\tilde{h}}
\def\ft{\tilde{f}}
\def\gt{\tilde{g}}
\def\pt{\tilde{p}}
\def\qt{\tilde{q}}
\def\vt{\tilde{v}}
\def\nt{\tilde{n}}
\def\ut{\tilde{u}}
\def\wt{\tilde{w}}
\def\zt{\tilde{z}}
\def\xt{\tilde{x}}
\def\yt{\tilde{y}}
\def\Psit{\tilde{\Psi}}
\def\vphit{\tilde{\varphi}}

\def\eb{\bar{\epsilon}}
\def\delb{\bar{\partial}}
\def\thb{\bar{\theta}}
\def\mub{\bar{\mu}}
\def\lamb{\bar{\l}}
\def\psib{\bar{\psi}}
\def\sb{\bar{\sigma}}
\def\xib{\bar{\xi}}
\def\chib{\bar{\chi}}

\def\Psib{\bar{\Psi}}
\def\Phib{\bar{\Phi}}
\def\Lamb{\bar{\Lambda}}
\def\Sb{{\overline \Sigma}}
\def\cb{\bar{c}}
\def\hb{\bar{h}}
\def\qb{\bar{q}}
\def\wb{\bar{w}}
\def\ub{\bar{u}}
\def\zb{{\bar{z}}}
\def\Hb{\bar{H}}
\def\Qb{{\bar Q}}
\def\Omegab{\overline{\Omega}}
\def\ob{\overline{\omega}}

\def\Ab{{\overline A}} \def\Bb{{\overline B}} \def\Cb{{\overline C}}
\def\Db{{\overline D}} \def\Eb{{\overline E}} \def\Fb{{\overline F}}
\def\Gb{{\overline G}}
\def\Ib{{\overline I}}
\def\Jb{{\overline J}} \def\Kb{{\overline K}} \def\Lb{{\overline L}}
\def\Mb{{\overline M}} \def\Nb{{\overline N}} \def\Ob{{\overline O}}
\def\Pb{{\overline P}}  \def\Rb{{\overline R}}
 \def\Tb{{\overline T}} \def\Ub{{\overline U}}
\def\Vb{{\overline V}} \def\Wb{{\overline W}} \def\Xb{{\overline X}}
\def\Yb{{\overline Y}} \def\Zb{{\overline Z}}

\def\fb{{\overline f}}
\def\gb{{\overline g}}
\def\mb{{\overline m}}
\def\lb{{\overline l}}
\def\yb{{\overline y}}

\def\ldel{{\overleftarrow{\del}}}
\def\rdel{{\overrightarrow{\del}}}
\def\ldeldel{{\overleftarrow{\del^2}}}
\def\rdeldel{{\overrightarrow{\del^2}}}
\def\ldelb{{\overleftarrow{\bar{\del}}}}
\def\rdelb{{\overrightarrow{\bar{\del}}}}

\def\ba{{\bf a}}
\def\bk{{\bf k}}
\def\bl{{\bf l}}
\def\bp{{\bf p}}
\def\bq{{\bf q}}
\def\br{{\bf r}}
\def\bt{{\bf t}}
\def\bu{{\bf u}}
\def\bv{{\bf v}}
\def\bx{{\bf x}}
\def\by{{\bf y}}
\def\bR{{\bf R}}
\def\bV{{\bf V}}

\def\bone{{\bf 1}}

\def\va{{\vec a}}
\def\vk{{\vec k}}
\def\vp{{\vec p}}
\def\vq{{\vec q}}
\def\vx{{\vec x}}
\def\vy{{\vec y}}
\def\vu{{\vec u}}
\def\vv{{\vec v}}

\def\vs{{\vec \sigma}}
\def\vtau{{\vec \tau}}

\newcommand{\ov}[1]{\overrightarrow{#1}}

\def\frA{\mathfrak{A}}
\def\frB{\mathfrak{B}}
\def\frC{\mathfrak{C}}
\def\frD{\mathfrak{D}}
\def\frE{\mathfrak{E}}
\def\frF{\mathfrak{F}}
\def\frG{\mathfrak{G}}
\def\frH{\mathfrak{H}}
\def\frM{\mathfrak{M}}
\def\frN{\mathfrak{N}}
\def\frR{\mathfrak{R}}
\def\frW{\mathfrak{W}}

\def\fra{\mathfrak{a}}
\def\frb{\mathfrak{b}}
\def\frf{\mathfrak{f}}
\def\frg{\mathfrak{g}}
\def\frh{\mathfrak{h}}
\def\frl{\mathfrak{l}}
\def\frs{\mathfrak{s}}
\def\fri{\mathfrak{i}}
\def\frj{\mathfrak{j}}

\def\ma{\mathfrak{a}}
\def\mg{\mathfrak{g}}
\def\mh{\mathfrak{h}}
\def\mR{\mathfrak{R}}
\def\mN{\mathfrak{N}}

\def\d{\delta}\def\D{\Delta}\def\ddt{\dot\delta}

\def\pa{\partial} \def\del{\partial}
\def\xx{\times}
\def\uno{\mbox{1 \kern-.59em {\rm l}}}

\def\trp{^{\top}}
\def\inv{^{-1}}
\def\dag{{^{\dagger}}}
\def\pr{^{\prime}}

\def\rar{\rightarrow}
\def\lar{\leftarrow}
\def\lrar{\leftrightarrow}

\newcommand{\0}{\,\!}      
\def\one{1\!\!1\,\,}
\def\im{\imath}
\def\jm{\jmath}

\newcommand{\tr}{\mbox{tr}}
\newcommand{\slsh}[1]{/ \!\!\!\! #1}

\def\vac{|0\rangle}
\def\lvac{\langle 0|}

\def\hlf{\frac{1}{2}}
\def\ove#1{\frac{1}{#1}}

\def\Box{\square}
\def\CC {\mathbb{C}}
\def\FF {\mathbb{F}}
\def\RR{\mathbb{R}}
\def\NN{\mathbb{N}}
\def\ZZ{\mathbb{Z}}
\def\bb#1{{\bf #1}}
\def\bcomment#1{}
\def\bfhat#1{{\bf \hat{#1}}}
\def\VEV#1{\left\langle #1\right\rangle}

\newcommand{\ex}[1]{{\rm e}^{#1}} \def\ii{{\rm i}}

\newcommand{\lrbrk}[1]{\left(#1\right)}
\newcommand{\lrsbrk}[1]{\left[#1\right]}
\newcommand{\sfrac}[2]{{\textstyle\frac{#1}{#2}}}

\def\stw{{\sqrt{2}}}

\def\rf {{\rm f}}
\def\ri {{\rm i}}
\def\rj {{\rm j}}
\def\rk {{\rm k}}
\def\rl {{\rm l}}
\def\rs {{\scriptscriptstyle \rm S}}
\def\rt {{\scriptscriptstyle \rm T}}

\def\rQ {{\scriptscriptstyle \rm \cQ}}
\def\rR {{\scriptscriptstyle \rm \cR}}

\def\cQb{{\cal \Qb}}
\def\cRb{{\cal \Rb}}
\def\cWb{{\cal \Wb}}

\def\fd {{\rm N}}
\def\afd {{\overline{\rm N}}}

\def \II {I\hspace{-.1em}I\hspace{.1em}}
\def \IIA {\mbox{\II A\hspace{.2em}}}
\def \IIB {\mbox{\II B\hspace{.2em}}}
\def \gs {g^s}
\def \ls {\lambda^s}

\def \I {{\cal I}}
\def \qs {q\hspace{-.53em}/\hspace{.15em}}
\def \ks {k\hspace{-.53em}/\hspace{.15em}}
\def \YM {{\mbox{\tiny YM}}}
\def \gym {g_{\YM}}

\def \Lc {\L_c}
\def\IR{\relax{\rm I\kern-.18em R}}
\def \id {{\bf 1}}

\def\cci{\ell}
\def\ccj{\ell'}

\def \thbb{\overline{\th\th}}
\newcommand \ol{\overline}
\def \lamb{\bar{\lambda}}
\def \vphi{\varphi}
\def \lambh{\hat{\bar{\lambda}}}
\def \lh{\hat{\lambda}}
\def \dd{\ddagger}

\def \ad {\dot{a}}
\def \bd {\dot{b}}
\def \cd {\dot{c}}
\def  \ddd {\dot{d}}
\def \ed {\dot{e}}
\def \fd {\dot{f}}
\def \Bh {\hat{B}}
\def \zm {{(0)}}
\def \nz {{(\text{KK})}}
\def \3{{(3)}}
\def \diag {\text{diag}}
\def \inm {{(m^{-1})}}

\def\eh{{\hat{e}}}
\def\fh{{\hat{f}}}
\renewcommand{\mh}{\hat{m}}
\def\theequation{\thesection.\arabic{equation}}

\def\adj{\text{adj}}
\def\co{\text{co}}

\author{Sheng-Lan Ko
\footnote{sheng-lan.ko@durham.ac.uk}$~^a$,
Dmitri Sorokin
\footnote{dmitri.sorokin@pd.infn.it}$~^b$
and Pichet Vanichchapongjaroen
\footnote{pichet.vanichchapongjaroen@durham.ac.uk}$~^a$
\\
\\
{\small $^a$ \it Centre for Particle Theory
and Department of Mathematical Sciences,}
\\
{\small \it Durham University, Durham, DH1 3LE, UK   }
\\
\\
{\small $^b$ \it INFN, Sezione di Padova, via F. Marzolo 8, 35131 Padova, Italia}
}

\title{\bf The M5--brane action revisited}

\date{}
\begin{document}
\maketitle

\abstract{We construct an alternative form of the M5--brane action in which the six--dimensional worldvolume is subject to a covariant split into 3+3 directions by a triplet of auxiliary fields. We consider the relation of this action to the original form of the M5--brane action and to a Nambu--Poisson 5--brane action based on the Bagger--Lambert--Gustavsson model with the gauge symmetry of volume preserving diffeomorphisms.}

\section{Introduction}
The construction of duality--symmetric actions has been an active topic of research since the 1970's \cite{Zwanziger:1970hk,Deser:1976iy}. It has recently seen a revival of interest in relation with the discussion of possible finiteness of $N=8, D=4$ supergravity \cite{Kallosh:2011qt,Bossard:2011ij,Carrasco:2011jv,Chemissany:2011yv,Pasti:2012wv,Roiban:2012gi,Kuzenko:2012ht,Ivanov:2012bq}, and in connection with attempts of making progress in understanding the non--Abelian (2,0) $6d$ superconformal gauge theory \cite{Witten:1995zh} on the worldvolume of $N$ coincident M5--branes \cite{Lambert:2010wm,Lambert:2010iw,Douglas:2010iu,Singh:2011id,Lambert:2011gb,Lambert:2012qy,Ho:2011ni,Huang:2012tu,Chu:2012um,Chu:2012rk,Chu:2013hja,Bonetti:2012fn,Bonetti:2012st,Singh:2012qr,Kim:2012tr,Fiorenza:2012tb,Saemann:2012uq,Saemann:2012kj,Palmer:2012ya,Samtleben:2011fj,Bandos:2013jva,Saemann:2013pca,Chu:2013joa}. For a single M5--brane the complete set of equations of motion was derived in \cite{Howe:1996yn} and considered in detail in \cite{Howe:1997fb} using the superembedding approach put forward in \cite{Sorokin:1989zi} (see \cite{Bandos:1995zw} and e.g. \cite{Sorokin:1999jx,Bandos:2009xy} for review and a detailed list of references). A complete M5--brane action was constructed in \cite{Bandos:1997ui,Aganagic:1997zq} as a result of a step--by--step generalization \cite{Schwarz:1993vs,Perry:1996mk,Pasti:1996vs,Schwarz:1997mc,Pasti:1997gx} of a self--dual action for a free chiral 2--form gauge field \cite{Henneaux:1987hz,Henneaux:1988gg}. It was then shown that the non--linear self--duality relation \cite{Howe:1997vn} and the complete set of the equations of motion \cite{Bandos:1997gm} derived from the M5--brane action are equivalent to the manifestly covariant equations obtained from superembedding.

It is well known that to lift the duality symmetry to the level of the action one should deal with the issue of space--time covariance of the theory. In the non--manifestly $SO(1,5)$ Lorentz--invariant construction of the $6d$ chiral 2--form action by Henneaux and Teitelboim \cite{Henneaux:1987hz,Henneaux:1988gg} only an $SO(5)$ or $SO(1,4)$ \cite{Perry:1996mk} subgroup of $SO(1,5)$ is manifest. The construction can be made space--time covariant (diffeomorphism invariant) by introducing into the action a normalized gradient of an auxiliary scalar field $a(x)$ \cite{Pasti:1995ii,Pasti:1995tn}, \cite{Pasti:1996vs}. The manifestly covariant formulation significantly simplifies the construction of the consistent couplings of the self-dual field action to gravity and other fields, and its non--linear deformations.  Different gauge fixings of the value of $a(x)$ using an associated local symmetry (or its dualization \cite{Maznytsia:1998xw}) results in different non--covariant forms of the self--dual action. On the other hand, the self-duality equations obtained from the action can be cast into the manifestly covariant form which does not contain the auxiliary field $a(x)$, thus the latter completely disappears on the mass--shell without imposing any gauge fixing condition.

With the advent of the Bagger--Lambert--Gustavsson (BLG) model \cite{Bagger:2006sk,Bagger:2007jr,Gustavsson:2007vu}, an alternative construction of a 5--brane action based on the BLG action with the
gauge symmetry of volume preserving diffeomorphisms was put forward in \cite{Ho:2008nn,Ho:2008ve} (see \cite{Ho:2009zt,Chen:2010br} for a review and references and \cite{Gustavsson:2009qd,Gustavsson:2011af} for a related work). The space--time and duality symmetries of this construction were analyzed in detail in \cite{Pasti:2009xc,Furuuchi:2010sp}. The equivalence of this model to the M5--brane description of \cite{Bandos:1997ui,Aganagic:1997zq} is still to be proved, though some steps have already been undertaken in \cite{Bandos:2008fr,Pasti:2009xc} and various checks via comparison of classical solutions on the both sides have been carried out (see \cite{Chen:2010br} and references therein).

The relation between the two actions is not obvious, first, since the original non--linear M5--brane action is of a Dirac--Born--Infeld type whose chiral 2--form gauge field transforms under the usual Abelian gauge transformations, while the action of \cite{Ho:2008ve} is a polynomial of up to six order in the fields and has a Nambu--Poisson 3--algebra structure associated with an un--conventional gauge invariance under volume preserving diffeomorphisms. In \cite{Ho:2008nn,Ho:2008ve} it was conjectured that the Nambu--Poisson (NP) M5--brane model is related to the
conventional description of the M5--brane in a constant $C_3$--field background through a transformation analogous to the
Seiberg--Witten map \cite{Seiberg:1999vs}. Such a map between the fields and gauge transformations of the two models was constructed in \cite{Chen:2010br}, however the relation between the two actions still remains to be established. The second reason which hampers the resolution of this issue is that in the NP M5--brane model the manifest $SO(1,5)$ $6d$ Lorentz symmetry is naturally broken by the presence of multiple M2--branes and the $C_3$--field to $SO(1,2)\times SO(3)$, which corresponds to a $3+3=6$ ``splitting" of the six dimensions of the $M5$--brane worldvolume. In the original M5--brane action, as we mentioned above, the six dimensions split into 1+5. In \cite{Pasti:2009xc} it was shown that, even when reduced to the second order in the fields, the two duality--symmetric actions are not equivalent off the mass shell, though both produce the same self-duality equation for the 2--form gauge field.

The M5--brane case exemplified the fact that the Lagrangian description of the self--dual fields and duality--symmetric fields in general is not unique (see also \cite{Belov:2006jd,Monnier:2011mv}), and various free (quadratic) duality--symmetric actions in D dimensions with different splittings of $D=p+q+r+...$ corresponding to various ways of breaking manifest space--time symmetry have been constructed \cite{Chen:2010jgb,Huang:2011np}. These different off--shell formulations may be useful for studying issues of the quantization of the self-dual fields in topologically non--trivial backgrounds \cite{Witten:1996hc,Dolan:1998qk,Henningson:1999dm,Witten:1999vg,Belov:2006jd,Monnier:2011mv,Sevrin:2013nca,Chen:2013gca}.

As far as the M5--brane is concerned, it is advisable for a better understanding of the relation between the original M5--brane descriptions and the NP 5--brane,  to see whether the quadratic self--dual action of \cite{Ho:2008nn} with ``3+3 splitting" can be extended to a full non--linear action which is invariant under the conventional gauge transformations of the gauge field and which would produce the same equations of motion as the ones obtained from the superembedding  \cite{Howe:1996yn} and the action of \cite{Bandos:1997ui,Aganagic:1997zq,Bandos:1997gm}. This is the main goal of this paper.

 Our strategy  to achieve this goal is as follows\footnote{For analogous procedures of getting  manifestly duality--symmetric non--linear actions see e.g. \cite{Perry:1996mk,Bossard:2011ij,Pasti:2012wv}.}. We will start with the covariant form \cite{Pasti:2009xc} of the quadratic self--dual action of \cite{Ho:2008nn} for a 2--form chiral gauge field in six--dimensions. In addition to the conventional invariance under the gauge transformations of the chiral field, the covariant action possesses two more local symmetries. One of them ensures that the auxiliary fields, which make the action covariant, are non--dynamical and another one guarantees that the self--duality condition on the field strength of the chiral field is the general solution of its equations of motion. We will add to this quadratic action a generic non--linear function of components of the chiral field strength and derive conditions on the form of this function imposed by the two local symmetries. It is known that the conditions obtained in this way may have more than one solution (see e.g. \cite{Gaillard:1981rj,Gibbons:1995cv,Perry:1996mk,Bossard:2011ij,Pasti:2012wv}), so to single out the solution which describes the M5--brane we will look for the one which is equivalent to the non--linear self--duality relation of the superembedding approach. More concretely, we will check that the non--linearly self--dual field strength of the superembedding formulation satisfies the condition imposed on the non--linear part of the self--dual action and, as a result, will derive an explicit form of the M5--brane action in which the $6d$ diffeomorphism invariance is subject to ``3+3 splitting".

 As is known from an extensive literature (see e.g. \cite{Gibbons:1995cv,Bossard:2011ij,Carrasco:2011jv,Chemissany:2011yv,Pasti:2012wv,Kuzenko:2012ht} and references therein), in general, the functionals of gauge--field strengths which determine non--linear self--duality conditions are constructed  order--by--order as perturbative series expansions in powers of the field strength and in general their explicit form is unknown except for the Born--Infeld--type actions and few other examples (see e.g. \cite{Hatsuda:1999ys,Ivanov:2012bq}). Our construction is a new example of an explicit (closed) form of the non--linearly self--dual action which differs from the canonical form of the Born--Infeld--type actions by additional terms and factors.

 The paper is organized as follows. In Section 2 we introduce main notation and conventions. In Section 3 we review the original action and present the structure of the novel action for the M5--brane. The derivation of the new action is explained in Section 4. In Section 5 we show that on--shall values of the two actions are equal and in Section 6 briefly discuss the dimensional reduction of the novel M5--brane action to that of the M2--brane. The results are summarized in Conclusion, where we also discuss open issues and possible directions of further research. In the Appendix we give more details of the check of the form of the new M5--brane action by comparing the self--duality relations which follow from the action with those obtained in the superembedding description of the M5--brane.

\setcounter{equation}0
\section{Notation and Conventions}
The $6d$ and the $D=11$ Minkowski metrics have the almost plus signature, $x^\mu$ ($\mu=0,1,\cdots, 5$) stand for the worldvolume coordinates of the M5--brane  which carries the chiral gauge field $B_2(x)=\frac 12 dx^\mu dx^\nu B_{\nu\mu}(x)$. The $D=11$ bulk superspace is parametrized by $Z^{\mathcal M}=(X^M,\theta)$, where $X^M$ are eleven bosonic coordinates and $\theta$ are 32 real fermionic coordinates. The geometry of the $D=11$ supergravity are described by  tangent--space vector supervielbeins $E^A(Z)=dZ^{\mathcal M}E_{\mathcal M}{}^A(Z)$ ($A=0,1,\cdots 10)$ and Majorana--spinor supervielbeins $E^\alpha(Z)=dZ^{\mathcal M}E_{\mathcal M}{}^\alpha(Z)$ ($\alpha=1,\cdots 32)$.

The vector supervielbein satisfies the following essential torsion constraint, which is required for proving the kappa--symmetry of the $M5$--brane action,
\be\label{T}
T^A=DE^A=dE^A+E^B\Omega_B{}^A=-iE^\alpha \Gamma^A_{\alpha\beta}E^\beta\,,
\ee
where $\Omega_B{}^A(Z)$ is the one--form spin connection in $D=11$, $\Gamma^A_{\alpha\beta}=\Gamma^A_{\beta\alpha}$ are real symmetric gamma--matrices and the external differential acts from the right.

The induced metric on the $M5$--brane worldvolume is constructed with the pull--backs of the vector supervielbeins $E^A(Z)$
\be\label{g6}
g_{\mu\nu}(x)=E_\mu^AE^B_\nu
\eta_{AB},
\qquad E_\mu^A=\partial_\mu Z^{\mathcal N}E_{\mathcal N}{}^A(Z(x)).
\ee

The $M5$--brane couples to the $D=11$ supergravity 3--form gauge superfield
$C_3(Z)=\frac 1{3!}
dZ^{\mathcal M_1}dZ^{\mathcal M_2}dZ^{\mathcal M_3}
C_{\mathcal M_3 \mathcal M_2 \mathcal M_1}$
and its $C_6(Z)$ dual, their field strengths are constrained as follows
\bea\label{F47}
dC_3&=&-\frac i2 E^AE^BE^\alpha E^\beta(\Gamma_{BA})_{\alpha\beta}+\frac 1{4!} E^AE^BE^CE^DF^{(4)}_{DCBA}(Z)\,,\nonumber\\
dC_6-C_3dC_3&=&\frac{2i}{5!} E^{A_1}\cdots E^{A_5}E^\alpha E^\beta(\Gamma_{A_5\cdots A_1})_{\alpha\beta}+\frac 1{7!} E^{A_1}\cdots E^{A_7}F^{(7)}_{A_7\cdots A_1}(Z)\,\\
F^{(7)\,A_1\cdots A_7}&=&\frac 1{4!}\epsilon^{A_1\cdots A_{11}}F^{(4)}_{A_8\cdots A_{11}}\,, \qquad \epsilon^{0...10}=-\epsilon_{0...10}=1.\nonumber
\eea
The generalized field strengths of $B_2(x)$ which appears in the M5--brane action is
\be\label{H3}
H_3=dB_2+C_3\,,
\ee
where $C_3(Z(x))$ is the pullback on the M5--brane worldvolume of the 3--form gauge field.

\setcounter{equation}0
\section{M5-brane actions}
We start by briefly reviewing the original form of the M5--brane action and then will present our main result, namely, the alternative worldvolume action for the M5--brane in a generic $D=11$ supergravity background.
\subsection{Original M5--brane action}
In this case to ensure the $6d$ worldvolume covariance of the $M5$--brane action one uses a normalized gradient of the auxiliary scalar field $a(x)$ which can be chosen to be time--like or space--like, e.g.
\be\label{v}
v_\mu(x)=\frac{\partial_\mu a}{\sqrt{\partial_\nu a \,g^{\nu\lambda}(x)\,\partial_\lambda a}}\,, \qquad v_\mu v^\mu=1
\ee
The both choices are equivalent since in the action $\partial_\mu a$ appears only in the projector of rank one
\be\label{P}
P_\mu{}^{\nu}(x)=\frac{\partial_\mu a\partial^\nu a}{(\partial a)^2},\qquad PP=P,\qquad (\partial a)^2\equiv \partial_\nu a \,g^{\nu\lambda}\,\partial_\lambda a=\partial_\nu a\,\partial^\nu a\,.
\ee
This projector singles out one worldvolume direction from the six, i.e. makes the 1+5 covariant splitting of the $6d$ worldvolume directions.

The $M5$--brane action in a generic $D=11$ supergravity superbackground constructed in \cite{Pasti:1997gx,Bandos:1997ui,Aganagic:1997zq} has the following form:
\bea \label{PSTM5}
S
&=& + 2\int_{\mathcal{M}_6} d^6x \lrsbrk{\sqrt{ - \det(g_{\m\n} + i\Ht_{\m\n})}  + \frac{\sqrt{-g}}{4(\del a)^2} \del_\l a \Ht^{\l\m\n}H_{\m\n\r}\del^\r a } \nn\\
&&
 - \int_{\mathcal{M}_6} \lrbrk{ C_6 + H_3 \wedge C_3 },
\eea
with
\be
\Ht^{\r\m\n} \equiv \frac{1}{6\sqrt{-g}} \, \e^{\r\m\n\l\s\t} H_{\l\s\t}, \quad \Ht_{\m\n} \equiv \frac{\partial^\r a}{\sqrt{(\del a)^2}} \Ht_{\r\m\n}, \quad g=\det g_{\m\n}\,,
\ee
where
$$
\epsilon^{0\cdots 5}=-\epsilon_{0\cdots 5}=1\,.
$$

In addition to the conventional abelian gauge symmetry for the chiral 2-form,
the action (\ref{PSTM5}) has also the following two local gauge symmetries :
\be \label{15SYM1}
\d B_{\mu\nu} = 2\del_{[\mu} a\, \Phi_{\nu]}(x), \qquad \d a(x) = 0,
\ee
as well as
\be \label{15SYM2}
\d a = \vphi(x),\qquad \d B_{\mu\nu} = \frac{\vphi(x)}{\sqrt{(\del a)^2}} ( H_{\mu\nu} - \cV_{\mu\nu}),
\ee
where
\be\label{calV}
\cV^{\mu\nu}(\tilde H)
\equiv -2\,\frac{\d \sqrt{\det(\delta^\nu_{\mu} + i\Ht_{\mu}{}^{\nu})}}{\d \Ht_{\mu\nu}},
\quad H_{\mu\nu} \equiv H_{\mu\nu\rho} \frac{\partial^\rho a(x)}{\sqrt{(\del a)^2}},
\ee
with $\vphi(x)$ and $\Phi_\mu(x)$ being arbitrary local functions on the woldvolume.
The first symmetry (\ref{15SYM1}) ensures that the equation of motion of $B_2$ reduces to the non--linear self--duality condition
\be\label{sdM5o}
H_{\mu\nu}= \cV_{\mu\nu}(\tilde H)\,,
\ee
while the second symmetry (\ref{15SYM2}) is responsible for the auxiliary nature of the scalar field $a(x)$ and the $6d$ covariance of the action.

The action \eqref{PSTM5} is also invariant under the local fermionic kappa--symmetry transformations with the parameter $\kappa^\alpha(x)$ which act on the pullbacks of the target--space supervilebeins and the $B_2$ field strength as follows
\bea\label{PSTkappa}
i_\k E^\alpha \equiv \delta_\k Z^{\mathcal M} E^\alpha_{\mathcal M} = \frac 12 (1+\bar\Gamma)^\alpha{}_{\beta} \k^{\beta}, \quad
i_\k E^A \equiv \delta_\k Z^{\mathcal M} E^A_{\mathcal M}  =  0. \\
\delta g_{\mu\nu} = -4i E^\alpha_{(\mu} (\G_{\nu)})_{\alpha\beta} \, i_\k E^{\beta}, \quad
\d{H}^\3 = i_\k d C^\3, \quad \delta_\k a(x) = 0\,,    \nn
\eea
where $(1+\bar\Gamma)/2$ is the projector of rank 16 with $\bar \Gamma$ having the following form
\bea\label{barGPST}
\!\!\!\!\!\!\!\!\!\sqrt{\det(\delta_{\mu}^{\nu} + i\tilde H_{\mu}{}^{\nu})} \,\bar\Gamma
&=& \gamma^{(6)}
- \frac{1}{2}\G^{\mu\nu\lambda} P_{\mu}{}^{\rho}\tilde H_{\nu\lambda\rho}
 - \ove{16\sqrt{-g}} {\e^{\mu_1\cdots\mu_6}}
		\tilde H_{\mu_1\mu_2\lambda}\tilde H_{\mu_3\mu_4\rho}P^{\lambda\rho} \G_{\mu_5\mu_6}, \nn\\
\bar\Gamma^2&=& 1\,,\qquad \tr{\bar\Gamma}=0,
\eea
where
\be\label{gamma6}
\Gamma_\mu=E_\mu{}^A\Gamma_A\,,\qquad \gamma^{(6)}=\frac 1{6!\sqrt{-g}}\epsilon^{\mu_1\cdots\mu_6}\Gamma_{\mu_1\cdots\mu_6}\,.
\ee

\subsection{New M5--brane action}
For this case, to ensure worldvolume covariance of the construction, instead of the single scalar field we need to introduce a triplet of auxiliary scalar fields $a^s(x)$ with the index
$(s =1,2,3)$ labeling a 3-dimensional representation of $GL(3)$ which is an \emph{internal} global symmetry of the action. The partial derivatives of the scalars are used to construct the projector matrices \cite{Pasti:2009xc}
\be\label{PPi}
P_\mu{}^\nu = \del_\mu a^r Y^{-1}_{rs} \del^\nu a^s, \qquad \P_\mu{}^\nu = \delta_\mu^\nu - P_\mu{}^\nu, \qquad \P_\mu{}^\nu\,\partial_\nu a=0
\ee
with $Y^{-1}_{rs}$ being the inverse matrix of
\be
Y^{rs} \equiv \del_\l a^r \del_\r a^s g^{\l\r} .
\ee
The projectors identically satisfy the following differential condition
\be\label{dP}
\Pi_{[\rho}{}^\lambda\Pi_{\kappa]}{}^\mu D_{\lambda}P_{\mu}^\nu=0=\Pi_{[\rho}{}^\lambda\Pi_{\kappa]}{}^\mu D_{\lambda}\Pi_{\mu}^\nu\,\,
\ee
where $D_\mu$ is the worldvolume covariant derivative with respect to the induced metric $g_{\mu\nu}$.

Note that the projectors \eqref{PPi} have rank 3 and thus effectively split the 6d directions into 3+3 ones orthogonal to each other.

The new M5--brane action coupled to a curved superbackground has the following form
\be \label{33M5U}
S = \int_{\mathcal{M}_6} d^6x\,
	\lrsbrk{ -\frac{\sqrt{-g}}{6} (\tilde G^{\m\n\r} G_{\m\n\r} + 3\tilde F^{\m\n\r}F_{\m\n\r}) + 2\mathcal L_{M5}(F,G) }
	- \int_{\mathcal{M}_6} \lrbrk{ C_6 + H \wedge C_3 },
\ee
where
\bea\label{LM5}
\mathcal L_{M5} &=& - \frac{1}{36(1+G^2)} \e^{\m_1\mu_2\mu_3\mu_4\mu_5\mu_6} G_{\m_1\m_2\m_3}F_{\m_4\nu\lambda} F_{\m_5}{}^{\lambda\k} F_{\m_6\k}{}^\nu     \nn\\
&& + \frac{1}{1+G^2} \sqrt{-\det\lrbrk{g_{\m\n} + \ove{2}(F+G)_{\m\r\s}(F+G)_\n{}^{\r\s}}})
\eea
and $F_{\m\n\r}$ and $G_{\m\n\r}$ are components of the field strength $H_{\m\n\r}$ projected as follows
\bea
F_{\m\n\r} \equiv H_{\t\s\l} P^\t_\m \P^\s_\n \P^\l_\r, && \!\!\!\! G_{\m\n\r} \equiv H_{\t\s\l} \P^\t_\m \P^\s_\n \P^\l_\r,\quad
G^2 \equiv \ove{6} H_{\m\n\r} \P^\m_\t \P^\n_\s \P^\r_\l H^{\t\s\l},\label{GF}\\
\Ft_{\m\n\r} \equiv \Ht_{\t\s\l} P^\t_\m \P^\s_\n \P^\l_\r,  && \Gt_{\m\n\r} \equiv \Ht_{\t\s\l} \P^\t_\m \P^\s_\n \P^\l_\r, \label{tildeGF}\,.
\eea

The action enjoys the following two local gauge symmetries analogous to eqs. \eqref{15SYM1} and \eqref{15SYM2}.
The first one is
\be \label{33SYM1}
\d B_{\mu\nu} = P^{\r}_\mu P^\s_{\nu} \Phi_{\r\s}(x),\qquad \d a^s = 0,
\ee
where $\Phi_{\r\s}(x)$ are arbitrary parameters. Note that in view of the conditions \eqref{dP} it follows that the projected field strengths \eqref{GF}, and hence $\mathcal L_{M5}(G,F)$, are invariant under this symmetry
\be\label{dGF}
\delta_\Phi G_{\mu\nu\rho}=\delta_\Phi F_{\mu\nu\rho}=0,
\ee
while their dual \eqref{tildeGF} are not.

The second symmetry ensures the triplet of the scalar fields $a^s(x)$ to be auxiliary
\be \label{33SYM2}
\d a^s = \vphi^s(x), \quad
\d B_{\m\n}
= \frac{1}{2}\vphi^r Y^{-1}_{rs} \del^\r a^s\,
	\e_{\m\n\r\t\s\l} \lrbrk{\sqrt{-g}\tilde F^{\t\s\l} - \frac{\del \mathcal L_{M5}}{\del F_{\t\s\l}}},
\ee
where $\vphi^s(x)$ are local parameters\footnote{In what follows we will use a normalization of the functional derivative, denoted by $\frac{\partial \mathcal L(F)}{\partial F_{\mu\nu ...}}$, which differs from the one defined in \eqref{calV}. Namely, by definition the variation of a p--form $F_{\mu_1\cdots \mu_p}$ and the corresponding functional derivatives are defined as follows $\delta F_{\mu_1\cdots \mu_p} =\delta F_{\nu_1\cdots \nu_p}\frac{\delta F_{\mu_1\cdots \mu_p}}{\delta F_{\nu_1\cdots \nu_p}}=\delta F_{\nu_1\cdots \nu_p}\,\frac 1{p!}\,\frac{\del F_{\mu_1\cdots \mu_p}}{ \del F_{\nu_1\cdots \nu_p}}$. So that
$$
\frac{\partial \mathcal L}{\partial F_{\nu_1\cdots \nu_p}}\equiv p!\,\frac{\delta \mathcal L}{\delta F_{\nu_1\cdots \nu_p}}.
$$}.

This symmetry allows one to gauge fix $a^s(x)$ to coincide with three world--sheet coordinates, e.g. $x^a$ $(a=0,1,2)$ or $x^i$ $(i=3,4,5)$,
thus getting a non--covariant but non--manifestly worldsheet diffeomorphism invariant M5--brane action. For instance, let us impose the gauge fixing condition
\be\label{agauge}
a^s=\delta^s_a\,x^a\,,
\ee
identifying $a^s$ with $x^a$. Then the following combination of the worldvolume diffeomorphism $\delta x^\mu=\xi^\mu(x)$ and the local symmetry \eqref{33SYM2} leaves this gauge condition intact
$$
\delta a^s(x)=\xi^\mu(x)\partial_\mu a^s+\varphi^s(x)=\xi^s(x)+\varphi^s(x)=0,\quad \rightarrow \quad \varphi^s(x)=-\xi^s(x).
$$
Under the local transformation combined of the $6d$ diffeomorphism $\delta x^\mu=\xi^\mu(x)$ and the local variation \eqref{33SYM2} with $\varphi^a(x)=-\xi^a(x)$ the gauge field $B_{\mu\nu}$ transforms as follows
$$
\Delta_{\xi^\mu} B_{\mu\nu}=\delta_{\xi^\mu} B_{\mu\nu}- \frac{1}{2}\xi^a(x) \,
	\e_{a\m\n\t\s\l} \lrbrk{\sqrt{-g}\tilde F^{\t\s\l} - \frac{\del \mathcal L_{M5}}{\del F_{\t\s\l}}}\,,
$$
while the other M5--brane fields $X^M(x)$ and $\theta^\alpha(x)$ being transformed in the conventional way as worldvolume scalars.
In the gauge \eqref{agauge} the action \eqref{33M5U}, \eqref{LM5} is non--manifestly invariant under the modified worldvolume diffeomorphisms of the above form.

Upon tedious computations we have checked that the action is invariant under the kappa--symmetry transformations \eqref{PSTkappa} but with a $\bar\Gamma$ projector which has the following form
\bea\label{barG33}
&&  \!\!\!\! \!\!\!\! \!\!\!\! \!\!\!\!\!\!\!\! \!\!\!\! \!\!\!\! \!\!\!\! \!\!\!\! \!\!\!\!\frac{1}{1+G^2} \sqrt{\det\lrbrk{\delta_{\m}^{\n} + \ove{2}(F+G)_{\m\r\s}(F+G)^{\n\r\s}}}\, \bar{\Gamma} =    \nn\\
&=& \gamma^{(6)} + \frac{1}{6} \gamma^{(6)} (3F+G)^{\mu\nu\rho} \G_{\mu\nu\rho}
	 + \ove{2(1+G^2)} \gamma^{(6)} F^{\m\n\t} F^{\r\l}{}_{\t} \G_{\m\n\r\l}   \nn\\
	&&
	+ \ove{6(1+G^2)} \gamma^{(6)} \G^{\m\n\r} \big( 3(FFG)_{\m\n\r} + (FFF)_{\m\n\r} \big),
\eea
where
\bea
(FFG)_{\m\n\r} \equiv F_\m{}^{\t\s}{} F_{\n\s\l} G_\rho{}^\l{}_{\t}, \quad (FFF)_{\m\n\r} \equiv F_\m{}^{\t\s} F_{\n\s\l} F_{\r}{}^\l{}_{\t}.
\eea
\if{}
In the two special cases in which either $F = 0$ or $G = 0$ the matrix $\bar\Gamma$ reduces to
\bea\label{bGnew}
\sqrt{1+G^2} \,\bar{\Gamma}
&=& \g^{(6)}(1 + \frac 16  \G^{\m_1\m_2\m_3} G_{\m_1\m_2\m_3}), \quad \text{for $F=0$},   \nn\\
\sqrt{\det(\d_\m^\n + \ove{2}F_{\m\r\s}F^{\n\r\s})} \,\bar{\Gamma}
&=&
 \g^{(6)}(1 + \frac12 \G^{\m_1\m_2\m_3} F_{\m_1\m_2\m_3}
	+ \frac 12 F_{\m_1\m_2\r} F_{\m_3\m_4}{}^{\r} \G^{\m_1\m_2\m_3\m_4}  \nn\\
&&  \!\!\!\!\!\!\!\!\!\! + \frac 1{3!} \G_{\m_1\m_2\m_3}
F^{\m_1\n_1\n_2} F^{\m_2}{}_{\n_2\n_3}  F^{\m_3\n_3}{}_{\n_1}),
\quad \text{for $G=0$}.
\eea
\fi
Note that the term multiplying $\bar\G$ on the left hand side of \eqref{barG33} is equal (modulo $\sqrt{-\det g_{\mu\nu}}$) to the last term of the non--linear part \eqref{LM5} of the M5--brane Lagrangian.

Finally, the non--linear self--duality condition which is obtained from action \eqref{33M5U} as the consequence of the equations of motion of $B_2$ (see eq. \eqref{B2eom} of the next Section) has the following form
\be\label{33sd}
\tilde G^{\m\n\r} =\ove{\sqrt{-g}} \lrbrk{\frac{\del \mathcal L_{M5}}{\del G}}^{\m\n\r}\,,\qquad \tilde F^{[\m\n\r]}
=  \frac{1}{\sqrt{-g}} \lrbrk{\frac{\del\mathcal L_{M5}}{\del F}}^{[\m\n\r]}.
\ee
As we will show, this self--duality condition is related to eq. \eqref{sdM5o} via the manifestly covariant self--duality relation which comes from the superembedding approach \cite{Howe:1996yn}.

\setcounter{equation}0
\section{Derivation of the new M5--brane action}
To get the new M5--brane action \eqref{33M5U} we start from the covariant form \cite{Pasti:2009xc} of the quadratic action \cite{Ho:2008nn} for the 6d chiral field. It is obtained from \eqref{33M5U} by truncating the latter to the second order in the chiral field strength $H_3$ \footnote{For simplicity, but without loss of generality, we consider (for a moment) the pullbacks of the $11D$ gauge fields be zero.}
\bea\label{2ndorder}
S& = &  \frac {1}6  \int d^6 x \, \sqrt{-\det g_{\m\n}}\, (H-\tilde H)_{\mu\nu\rho}(\Pi_{\lambda}{}^\mu\Pi_\kappa{}^\nu\Pi_\tau{}^\rho+3\Pi_{\lambda}{}^\mu\Pi_\kappa{}^\nu P_\tau{}^\rho)H^{\lambda\kappa\tau}\nonumber\\
&\equiv & \frac {1}6  \int d^6 x \, \sqrt{-\det g_{\m\n}}\,  [(G-\tilde G)_{\mu\nu\rho}G^{\mu\nu\rho}+3(F-\tilde F)_{\mu\nu\rho}F^{\mu\nu\rho}]\,,
\eea
The action is invariant under the symmetry \eqref{33SYM1} and under the linearized counterpart of \eqref{33SYM2}
\be \label{33SYM2l}
\d a^s = \vphi^s(x), \quad
\d B_{\m\n}
= \frac{1}{2}\vphi^s Y^{-1}_{sr} \del^\r a^r\,
	\e_{\m\n\r\t\s\l}  \sqrt{-g}\lrbrk{\tilde F^{\t\s\l} - F^{\t\s\l}}.
\ee
The quadratic action leads to the equation of motion
\be
\del_\r \lrbrk{ \sqrt{-g} (G-\Gt)^{\m\n\r} + 3\sqrt{-g}(F-\Ft)^{[\m\n\r]} } = 0,
\ee
which has the general solution
\be
\sqrt{-g} (G-\Gt)^{\m\n\r} + 3 \sqrt{-g} (F-\Ft)^{[\m\n\r]}
= \ove{2}\e^{\m\n\r\t\s\l} \del_\t \lrsbrk{ \sqrt{-g} \tilde{\Phi}_{\eta\xi}P^\eta_\s P^\xi_\l },
\ee
with an arbitrary tensorial function $\tilde{\Phi}_{\eta\xi}$.
This function can be compensated by a gauge transformation
of the equation of motion under \eqref{33SYM1}
with the gauge parameter $-\tilde{\Phi}_{\xi\eta}$.
Hence, in view of the definition of the projected components of the field strength \eqref{GF}, the solution of the dynamical equation is equivalent to the self-duality conditions
\be
(G-\Gt)^{\m\n\r} = 0,\quad (F-\Ft)^{[\m\n\r]} = 0.
\ee

We are now looking for a non--linear generalization of the action \eqref{2ndorder} which would respect the both symmetries \eqref{33SYM1} and \eqref{33SYM2}. Note that the second symmetry should be deformed by the non--linear terms, since the form of its transformation is associated with the form of the non--linear self--duality condition. In the case of the M5--brane these are \eqref{15SYM2}--\eqref{sdM5o}, and \eqref{33SYM2} and \eqref{33sd}.

Since the field strength components $F_{\mu\nu\rho}$ and $G_{\mu\nu\rho}$ are invariant under the transformations \eqref{33SYM1} (see eqs. \eqref{dGF}), while their dual \eqref{tildeGF} are not, the non--linear terms in the action should only depend on $F$ and $G$. So the general form of the non--linear action which  respects the symmetry \eqref{33SYM1} is obtained by replacing the quadratic terms $FF$ and $GG$ in \eqref{2ndorder} by an arbitrary function $\mathcal L(F,G)$
\be \label{generic}
S = \int_{\mathcal{M}_6} d^6x\,
	\lrbrk{ -\frac{\sqrt{-g}}{6} (\tilde G^{\m\n\r} G_{\m\n\r} + 3\tilde F^{\m\n\r}F_{\m\n\r}) + 2\mathcal L(F,G) }.
\ee
The variation of this action with respect to the gauge potential $B_2$ produces the equations of motion
\be\label{B2eom}
\del_\rho \lrsbrk{ \lrbrk{\frac{\del\cL}{\del G}}^{\m\n\r} - \sqrt{-g} \Gt^{\m\n\r} + 3\lrbrk{\frac{\del\cL}{\del F}}^{[\m\n\r]} -
3 \sqrt{-g} \Ft^{[\m\n\r]} } = 0.
\ee
In view of \eqref{dGF} and the fact that $\cL$ only depends on $F$ and $G$,
we can integrate the above equation of motion with the help of the symmetry \eqref{33SYM1}
along the same lines as in free theory.
The integration produces the non--linear self--duality relations
\be\label{33sdg}
\tilde G^{\m\n\r} = \frac{1}{\sqrt{-g}} \lrbrk{\frac{\del \mathcal L}{\del G}}^{\m\n\r}\,,\qquad \tilde F^{[\m\n\r]}
=  \frac{1}{\sqrt{-g}} \lrbrk{\frac{\del\mathcal L}{\del F}}^{[\m\n\r]}.
\ee
We should now find conditions on the form of $\mathcal L(F,G)$ imposed by the requirement that the action is invariant under
\be \label{33SYM2g}
\d a^s = \vphi^s(x), \quad
\d B_{\m\n}
= \frac{1}{2}\vphi^s Y^{-1}_{sr} \del^\r a^r\,
	\e_{\m\n\r\t\s\l} \lrbrk{ \sqrt{-g} \tilde F^{\t\s\l} - \frac{\del \mathcal L}{\del F_{\t\s\l}}}.
\ee
Upon somewhat lengthy calculations using, in particular, the properties of the projectors \eqref{PPi}--\eqref{dP} and the form of their variation under \eqref{33SYM2g}
\be\label{deltaP}
\delta_\varphi P_{\mu\nu}= 2\Pi_{\rho(\mu}\partial^\rho \varphi^r Y^{-1}_{rs}\partial_{\nu)}a^s
\ee
we get the following condition on  $\mathcal L(F,G)$
\bea\label{nonlsd}
&&\!\!\!\!\!\!\!\!\!\!\del_{\mu} \left[
 Y^{-1}_{rs}\del^\nu a^s
 \left( \sqrt{-g} \lrbrk{\frac{\del\cL}{\del G}}^{\m\t\s} F_{\n\t\s} - \sqrt{-g} G^{\m\t\s}\lrbrk{\frac{\del\cL}{\del F}}_{\n\t\s}   \right.\right. \nn\\
&&\left.\left. \qquad\qquad
- \frac{g}{2}\e_{\n\t\s\l\xi\eta} F^{\l\xi\eta} F^{\t\s\m}
- \ove{2}\e_{\n\t\s\l\xi\eta} \lrbrk{\frac{\del\cL}{\del F}}^{\l\xi\eta}\lrbrk{\frac{\del\cL}{\del F}}^{\t\s\m}
\right)\right]  = 0.
\eea
This condition is analogous to those found in other instances of models with non--linear (twisted) self--duality, e.g in $D=6$ \cite{Perry:1996mk} and $D=4$ \cite{Bossard:2011ij,Pasti:2012wv}. It is well known that these conditions may have different solutions leading to different non--linear generalizations of quadratic duality--symmetric actions (see e.g. \cite{Gaillard:1981rj,Gibbons:1995cv,Perry:1996mk,Bossard:2011ij,Pasti:2012wv}). We are interested in a particular solution of the above equation, i.e. in the form of $\mathcal L(F,G)$ which describes the $M5$--brane. To find this form we assume that, as in the case of the self--duality condition \eqref{sdM5o} obtained from the original M5--brane action, also the self--duality conditions \eqref{33sd} (or \eqref{33sdg}) should be equivalent to the self--duality conditions appearing in the superembedding formulation of the M5--brane \cite{Howe:1996yn}. Exploring these conditions we shall derive the form \eqref{LM5} of the non--linear M5--brane Lagrangian.

\subsection{Non--linear self--duality of the M5--brane in the superembedding approach}
In the superembedding description of the $M5$--brane \cite{Howe:1996yn,Howe:1997fb} the field strength $H_3$ of the chiral field $B_2$ is expressed in terms of a  self--dual tensor $h_3=*h_3$ as follows\footnote{Our normalization of the field strength differs from that in \cite{Howe:1997vn} by the factor of $\frac 14$ in front of $H_3$.}
\be\label{embsd}
\frac 14 H_{\m\n\rho}=m^{-1\lambda}_{\mu}h_{\lambda\n\rho}\,, \qquad \frac 14\tilde H^{\m_1\n_1\rho_1}=\frac{1}6\epsilon^{\m_1\n_1\rho_1\m\n\rho}m^{-1\lambda}_{\mu}h_{\lambda\n\rho}=Q^{-1}m^{\mu_1\lambda}
h_{\lambda}{}^{\n_1\rho_1}\,
\ee
where $m^{-1\lambda}_{\mu}$ is the inverse matrix of
\be\label{mk}
m_{\mu}{}^{\lambda}=\delta_{\mu}{}^{\lambda}-2k_{\mu}{}^{\lambda}\,,\qquad m_\mu^{-1\lambda}=Q^{-1}(2\delta_{\mu}{}^{\lambda}-m_{\mu}{}^{\lambda}),\qquad k_{\mu}{}^{\lambda}=h_{\mu\nu\rho}h^{\lambda\nu\rho}\,
\ee
and
\be\label{Q}
Q=1-\frac 23 \tr \,k^2\,.
\ee
As was shown in \cite{Howe:1997vn}, by splitting the indices in eqs. \eqref{embsd} into 1+5 and expressing components of $h_3$ in terms of $\tilde H_{\mu\nu 5}$ one gets the duality relation \eqref{sdM5o}\footnote{This splitting is amount to projecting the tensor fields along the direction of ${\partial_\mu a}$ and orthogonal to it.}.

We shall now carry out a similar procedure, but splitting the 6d indices into 3+3, and upon a somewhat lengthy algebra will arrive at the self--duality condition in the form of \eqref{33sd}, thus getting the non--linear function $\mathcal L_{M5}(F,G)$ \eqref{LM5} which enters the M5--brane action \eqref{33M5U}.

The 3+3 splitting can be performed with the use of the projectors \eqref{PPi}, but for computational purposes we have found it more convenient to pass to a local tangent--space frame using $6d$ vielbeins $e^m_\mu$ $(e^m_\m \eta_{mn}e^n_\nu=g_{\mu\nu})$
and to write the 3+3 tangent space indices explicitly. So the three directions singled out by the projector $P_m{}^n\equiv e^\mu_m P_\mu{}^{\nu}e_\nu^n$, which we assume to contain the time direction, will be labeled by the indices $a,b,c$, and the three spacial directions singled out by $\Pi_{m}{}^n\equiv  e^\mu_m \Pi_\mu{}^{\nu}e_\nu^n $
will be labeled by $i,j,k$:
\be\label{3+3}
P_m{}^n \rightarrow \delta_a^b\,,\qquad \Pi_{m}{}^n\rightarrow \delta_i^j\,, \qquad a,b,c=0,1,2;\qquad i,j,k=3,4,5\,,
\ee
while the $6d$ Levi--Civita tensor splits as follows
\be\label{LC}
\epsilon^{\mu\nu\rho\lambda\tau\kappa}\,\,\Rightarrow \epsilon^{abc}\epsilon^{ijk}, \qquad \epsilon^{012}=-\epsilon_{012}=1,\qquad \epsilon^{345}=1\,.
\ee
We are now ready to split the indices of $H_3$ and $h_3$ in \eqref{embsd}.
\subsubsection{3+3 splitting}
As $h_3$ is self--dual,
we pick its $10$ independent components in the local Lorentz frame as follows
\be
h_{ija},\quad h_{ijk}
\ee
and define\footnote{One should not confuse the field $g(x)$ with the determinant of the induced metric $g_{\mu\nu}$.}
\be
f_{a}^k \equiv \hlf\e^{ijk}h_{ija},
\quad
g \equiv \ove{6}\e^{ijk}h_{ijk}.
\ee
In view of the self--duality
\be
h^{mnp} = \ove{3!}\e^{mnpl_1l_2l_3} h_{l_1l_2l_3},
\ee
we have
\be
h^{jab} = -\e^{abc}f_c^j,
\quad h^{abc} = g\e^{abc},
\ee
or
\be
f_{ic} = \hlf\e_{abc}
	h_i{}^{ab}\,
\qquad g = -\ove{6}\e_{abc}h^{abc}.
\ee
The corresponding components of $H_3$ are defined as
\be\label{FG}
F_{a}^k \equiv \hlf\e^{ijk}H_{ija},
\quad
G \equiv \ove{6}\e^{ijk}H_{ijk}.
\ee
The duals of $F$ and $G$ are
\be\label{tildeFG}
\Ft_{ic} \equiv \hlf\e_{abc}H_i{}^{ab},
\quad \Gt \equiv -\ove{6}\e_{abc}H^{abc}.
\ee
Note that the tensors \eqref{FG} and \eqref{tildeFG} are counterparts of \eqref{GF} and \eqref{tildeGF} in the local Lorentz frame \eqref{3+3}.

Our final goal is to write $\Ft, \Gt$
in terms of $F,G$ using the relations \eqref{embsd}. To this end, using \eqref{embsd} we first find the expressions for $F,G,\tilde F$ and $\tilde G$ in terms of $g$ and $f_i^a$
\bea\label{F}
\frac 14 F_i^a &=& Q^{-1}\lrbrk{f(1+4g^2-4\tr f^2) + 8f^3 - 8gf^{-1} \det f}_i^a \nonumber\\
&=& Q^{-1}\frac{\pa}{\pa f_a^i}\lrbrk{\hlf(g^2+\tr f^2)-\ove{16}Q},
\eea
\bea\label{G}
\frac 14 G &=& Q^{-1}\lrbrk{g + 4g^3 + 4g\tr f^2 - 8  \det f}\nonumber\\
 &=& Q^{-1}\frac{\pa}{\pa g}\lrbrk{\hlf(g^2+\tr f^2)-\ove{16}Q},
\eea
and
\bea\label{tF}
\frac 14 \Ft_i^a &=& Q^{-1}\lrbrk{f(1-4g^2+4\tr f^2) - 8f^3 + 8g f^{-1} \det f}_i^a \nonumber\\
&= & Q^{-1}\frac{\pa}{\pa f_a^i}\lrbrk{\hlf(g^2+\tr f^2)+\ove{16}Q},
\eea
\bea\label{tG}
\frac 14 \Gt &= & Q^{-1}\lrbrk{g - 4g^3 - 4g\tr f^2 + 8  \det f}\nonumber\\
 &=& Q^{-1}\frac{\pa}{\pa g}\lrbrk{\hlf(g^2+\tr f^2)+\ove{16}Q},
\eea
where
\be\label{Qgf}
Q = 1-16g^4+16(\tr f^2)^2 - 32g^2\tr f^2 -32 \tr f^4 +128 \,g \det f,
\ee
\be
\tr f^2 \equiv f^a_if_j^b \delta^{ij} \eta_{ab}\,,\qquad \det f \equiv \ove{6}\e_{ijk}\e^{abc}f_a^i f_b^j f_c^k\,,\qquad  (f^{-1})^a_i \det f\equiv  \frac 12 \e_{ijk}\e^{abc}f_b^j f_c^k\,.
\ee
\subsubsection{The M5--brane action in terms of $G$ and $F^{i}_a$}
For the fields \eqref{FG} and \eqref{tildeFG} the M5--brane action \eqref{33M5U} takes the following form
\bea \label{eq:33actionScaled}
S_{3+3} = -\int d^6x\lrbrk{ \sqrt{-\det g_{\mu\nu}}(F_a^i\Ft_i^a +G\Gt)  - 2\mathcal L_{M5} }- \int_{\mathcal{M}_6} \lrbrk{ C_6 + H \wedge C_3 },
\eea
where the term $\mathcal L_{M5}$ is
\bea \label{LM5fixed}
\mathcal L_{M5} = {\sqrt{-\det g_{\mu\nu}}}\lrbrk{\frac{   G \det F }{1+G^2}
				  + \frac{\sqrt{\det \lrbrk{ \delta_{i}^{j}(1+G^2)+F_i^aF^j_{a} }}}{1+G^2}},
\eea
and the non--linear self--duality relations \eqref{33sd} become
\be\label{33sdfixed}
\tilde G=\frac 1{\sqrt{-\det g_{\mu\nu}}}\frac{\del \mathcal L_{M5}}{\del G}\,,\qquad \tilde F^{a}_i
= \frac 1{\sqrt{-\det g_{\mu\nu}}}\frac{\del\mathcal L_{M5}}{\del F_a^i}.
\ee

\subsubsection{Self--duality relations in particular cases}\label{cases}
To guess the form  \eqref{LM5fixed} of the function $\mathcal L_{M5}$  in the M5-brane action we first consider a number of simple cases.
\\
\\
\textbf{{$f=0$ case} }
\\
\\
The relations \eqref{F}-\eqref{Qgf} reduce to
$$
F_i{}^a=\tilde F_i{}^a=0\,,\qquad Q = 1-16g^4,
$$
\bea\label{Gg}
\frac 14 G &=&\frac{g + 4g^3}{1-16g^4}=\frac g{1-4g^2},
\eea
\bea\label{tGg}
\frac 14 \Gt &= & \frac{g - 4g^3}{1-16g^4}=\frac g{1+4g^2}.
\eea
We now solve eq. \eqref{Gg} for $g$
\be\label{g(G)}
g=\frac{\pm\sqrt{1+G^2}-1}{2G}\,.
\ee
Since, due to \eqref{Gg}, in the linear approximation $G/4=g$,
we should pick up only the solution with the upper sign.
Substituting this solution into \eqref{tGg} we get the relation between $\tilde G$ and $G$
\be\label{tG(G)}
\tilde G=\frac{G}{\sqrt{1+G^2}}=\frac{\partial\sqrt{1+G^2}}{\partial G}\,.
\ee
We see that eq. \eqref{tG(G)} is exactly the same as \eqref{33sdfixed} when in \eqref{LM5fixed} we put $F_{i}^a=0=F_{\mu\nu\rho}$. This demonstrates how the (square root of) factor $1+G^2$ appears in the function $\mathcal L_{M5}(F,G)$   \eqref{LM5} or  \eqref{LM5fixed} of the M5--brane action \eqref{33M5U}.
\\
\\
\textbf{{$g=\det f=0$ case} }
\\
\\
Now the relations  \eqref{F}--\eqref{Qgf} reduce to
$$
G=\tilde G=0\,,\qquad Q = 1+16(\tr f^2)^2 -32 \tr f^4,
$$
\bea\label{Ff}
\frac 14 F_i^a &=& Q^{-1}\lrbrk{f(1-4\tr f^2) + 8f^3 }_i^a
=Q^{-1}\frac{\pa}{\pa f_a^i}\lrbrk{\hlf\tr f^2-\ove{16}Q},
\eea
\bea\label{tFf}
\frac 14 \Ft_i^a &=& Q^{-1}\lrbrk{f(1+4\tr f^2) - 8f^3}_i^a
= Q^{-1}\frac{\pa}{\pa f_a^i}\lrbrk{\hlf\tr f^2+\ove{16}Q},
\eea
Let us simplify things even further by considering a solution of the non--linear self--duality equation such that the only non--zero components of $f^a_i$ are $f^1_i$. Then the above equations further reduce to
$$
G=\tilde G=0\,,\qquad Q = 1-16(f^2)^2, \qquad f^2\equiv f^1_if^1_i\,,
$$
\bea\label{Ff1}
\frac 14 F_i^1 &=& Q^{-1}\lrbrk{1+4 f^2 }f_i^1
=\frac{f_i^1}{1-4 f^2 },
\eea
\bea\label{tFf1}
\frac 14 \Ft_i^1 &=& \frac{f_i^1}{1+4 f^2 },
\eea
From these equations we find that
$$
1-4f^2=-\frac{2(1\mp \sqrt{1+F^2})}{F^2}\,,\qquad 1+4f^2=\frac{2\sqrt{1+F^2}}{F^2}(\sqrt{1+F^2}\mp 1)\,,
$$
$$
f_i^1=\frac{F^1_i}{2F^2}(\pm\sqrt{1+F^2}-1).
$$
Since, due to \eqref{Ff1}, in the linear approximation $F_i^a/4=f^a_i$,
in the above relation we should pick the upper sign and upon substituting it into \eqref{tFf} we get the duality relation
\be\label{tildeF1}
\tilde F^1_i=\frac {F_i^1}{\sqrt{1+F^2}}=\frac{\partial\sqrt{1+F^2}}{\partial F^i_1}\,.
\ee
We see that this relation coincides with  \eqref{33sdfixed} for $G=0$ and $F^a_i$ having only the non--zero components $F^1_i$.

\subsubsection*{Self--dual string soliton ({$g\not =0, \det f=0$ case}) }
Let us now consider a more complicated particular case of a string soliton solution of \cite{Perry:1996mk}. A similar consideration is applicable to the BPS self--dual string of \cite{Howe:1997ue}.
For the string aligned along the $x^2$--coordinate, in terms of fields (\ref{FG}) and (\ref{tildeFG}) the string soliton solution of \cite{Perry:1996mk} has the following form:
\be\label{Gss}
G = -\frac{\b x^1}{\r^4},\qquad
F_1^k = -\frac{\b x^k}{\r^4},
\ee
\be\label{tGss}
\Gt = -\frac{\a' x^1}{\r},\qquad
\Ft_1^k = -\frac{\a' x^k}{\r}.
\ee
where $k=3,4,5$, $\r:=\sqrt{x_1^2+x_3^2+x_4^2+x_5^2}$, $\b$ is a constant and
\be
\a'(\r) = \frac{\b}{\sqrt{\b^2+\r^6}}.
\ee
In this form the string soliton solution was considered in \cite{Chu:2012rk}. It naturally splits the $6d$ worldvolume into 3+3 directions.

The form  \eqref{Gss} of $G$ and $F$ suggests that in \eqref{F} and \eqref{G} $g\not =0$ and the non--zero components of $f_i^a$ are $f^1_i$. So the equations \eqref{F}--\eqref{Qgf} reduce to
\be\label{Qgfsd}
Q = 1-16(g^2+f^2)^2.
\ee
\be\label{FGsd}
\frac 14 F_i^1 = \frac{f_i^1}{1-4(g^2+  f^2)}\,,\qquad
\frac 14 G = \frac{g}{1-4(g^2+  f^2)},
\ee
and
\be\label{tFsd}
\frac 14 \Ft_i^1 = \frac{f_i^1}{1+4(g^2+  f^2)}\,,\qquad
\frac 14 \Gt =  \frac{g}{1+4(g^2+  f^2)}.
\ee
Carrying out the same analysis as in the previous examples, from \eqref{Qgfsd}--\eqref{tFsd} we get the duality relations
\be\label{tFF}
\tilde F_i^1=\frac{F_i^1}{\sqrt{1+G^2+F^2}}=\frac{\partial\sqrt{1+G^2+F^2}}{\partial F^i_1}\,,\qquad
\tilde G=\frac{G}{\sqrt{1+G^2+F^2}}=\frac{\partial\sqrt{1+G^2+F^2}}{\partial G}\,
\ee
which are again a particular case of \eqref{33sdfixed}. One can then guess that in the manifestly covariant formulation the expression under the square root combines into the determinant of the matrix formed by the bilinear combinations of $G_{\mu\nu\rho}$ and $F_{\mu\nu\rho}$ as in eq. \eqref{LM5} or \eqref{LM5fixed}.

To see that this is indeed so and that \eqref{LM5fixed} should also contain the term $G \det F$ let us consider the case in which $G=0$ while $F_i^a$ is (otherwise) generic.

\subsubsection*{$G =0$ case}
We have
\be\label{G0}
G =0=\lrbrk{g + 4g^3+ 4g\tr f^2 - 8  \det f}\,,
\ee
\be\label{Gt0}
\frac 14 \Gt  =   2Q^{-1} g\,,
\ee
\bea\label{Qtr}
Q &=& 1-16g^4+16(\tr f^2)^2 -32g^2\tr f^2-32 \tr f^4+ 128g \det f\nonumber\\
&=&1+16g^2+16(\tr f^2)^2+48g^4+32g^2\tr f^2-32 \tr f^4,
\eea
\be\label{F0}
\frac 14 F_i^a = Q^{-1}\lrbrk{f(1+4g^2-4\tr f^2) + 8f^3 - 8gf^{-1} \det f}_i^a\,
\ee
and
\bea\label{tF0}
\frac 14 \Ft_i^a = Q^{-1}\lrbrk{f(1-4g^2+4\tr f^2) - 8f^3 + 8g f^{-1} \det f}_i^a .
\eea
Now, the direct computation of $\det F$ using \eqref{F0} and \eqref{G0} gives (see also eq. \eqref{detF} of the Appendix)
\be\label{DetF}
\det F=8 Q^{-1} g\,.
\ee
Comparing this equation with \eqref{Gt0} we get
\be\label{GtF}
\Gt =\det F\,,
\ee
which is exactly the relation that we get by varying the term \eqref{LM5fixed} of the M5-brane action \eqref{33M5U} or \eqref{eq:33actionScaled} with respect to $G$ and setting $G=0$ afterwards. This explains the appearance of the term $G \det F$ in the M5--brane action.

On the other hand, upon expressing the right--hand side of \eqref{tF0} in terms of $F^a_{i}$ and performing somewhat lengthy computations using Mathematica one gets the duality relation for $\tilde F$ which coincides with eq. \eqref{33sdfixed} evaluated at $G=0$.

Finally, by a direct check using Mathematica one can verify that also in the generic case the components $F$, $\tilde F$, $G$ and $\tilde G$ of the field strength $H_3$ determined by the superembedding relations \eqref{F}--\eqref{Qgf} satisfy the non--linear duality relations \eqref{33sdfixed} which follow from the M5--brane action \eqref{33M5U}. Main steps of the calculation are described in the Appendix.

The last point that one should check is that the function \eqref{LM5} satisfies eq. \eqref{nonlsd} which insures the invariance of the M5--brane action under the local transformations \eqref{33SYM2}. The direct calculation shows that this is indeed so. Actually,  \eqref{LM5}
satisfies even stronger relation, namely, it makes to vanish the expression under the derivative in \eqref{nonlsd}.

\setcounter{equation}0
\section{Comparison of the two M5--brane actions}
As was discussed in \cite{Pasti:2009xc} duality symmetric actions corresponding to different splittings of space--time  differ from each other by terms that vanish on--shell, {\it i.e.} when (an appropriate part of) the self--duality relations is satisfied. In \cite{Pasti:2009xc} this was discussed for the free chiral 2--form in $6d$.

We shall now confront the two M5--brane actions \eqref{PSTM5} and \eqref{33M5U} by comparing their values for the 3-form field strength satisfying the non--linear self--duality equation. As we have seen, the non--linear self--duality relations that follow from these actions are similar and are equivalent to the self--duality condition that follows from the superembedding formulation. Therefore, to compute the on--shell values of the M5--brane actions we will substitute into them the expressions of the components of $H_3$ and $\tilde H_3$ in terms of the components of the self--dual tensor $h_3$.

In the case of the novel action these are eqs. \eqref{F}--\eqref{tG}. Substituting them into the action \eqref{33M5U} (or \eqref{eq:33actionScaled}) and using Mathematica we find that the on--shell value of the  self--dual M5--brane action is
\be\label{onshellM5}
S^{\text{on-shell}}_{M5}= 4\int d^6x \sqrt{-\det g_{\mu\nu}}\,\,Q^{-1}
 - \int_{\mathcal{M}_6} \lrbrk{ C_6 + H \wedge C_3}.
\ee
Notice that the
Lagrangian of this action
is the functional of $Q(h)$ defined in \eqref{Q}. We thus see that the on--shell action is manifestly $6d$  covariant and does not depend on the auxiliary fields $a^r(x)$ \eqref{PPi}.

To compute the corresponding on--shell value of the original M5--brane action \eqref{PSTM5} we perform the 1+5 splitting of the duality relations \eqref{embsd} which take the following form
$$
\Ht_{\ah\bh 5} = 4Q^{-1}\lrbrk{ (1-2\tr \fb^2)\fb + 8\fb^3 }_{\ah\bh 5}, \quad
H _{\ah\bh 5}= 4Q^{-1}\lrbrk{ (1+2\tr \fb^2)\fb - 8\fb^3 }_{\ah\bh 5},
$$
where $\fb_{\ah\bh}=h_{\ah\bh 5}$ and $\ah,\bh=0,1,2,3,4$. Upon substituting the above expressions into the action \eqref{PSTM5} we find that its value is again given by eq. \eqref{onshellM5}. Thus the two forms of the M5--brane action give rise to the same equations of motion and their on--shell values are equal and are given by the superembedding scalar function $Q(h)$. For the self--dual string soliton considered in Section \ref{cases}, the value of the action determines the tension of the string, as was discussed in \cite{Perry:1996mk}.

An interesting open problem that may have important consequences for the issue of quantization of the self--dual fields is the understanding of the off--shell relationship between the different self--dual actions.

\section{Relation to $M2$--branes}
The new form of the $M5$--brane action can be useful for studying its relation to the Nambu--Poisson description of the M5--brane in a constant $C_3$ field originated from the $3d$ BLG model with the gauge group of volume preserving diffeomoprhisms \cite{Ho:2008nn,Ho:2008ve}. The BLG model invariant under the volume preserving diffeomorphisms describes a condensate of M2--branes which via a Myers effect may grow into an M5--brane. In \cite{Ho:2008nn,Ho:2008ve} it was conjectured that the Nambu--Poisson M5--brane model is related to the
conventional description of the M5--brane in a constant C--field background through a transformation analogous to the
Seiberg--Witten map \cite{Seiberg:1999vs}. Such a map between the fields and gauge transformations of the two models was constructed in \cite{Chen:2010br}, however the relation between the two actions still remains to be established. We leave the study of this issue for future and will only show that in a flat background without C--field the worldvolume dimensional reduction of the bosonic M5--brane action \eqref{33M5U} (or \eqref{eq:33actionScaled}) directly results in the membrane action. To this end we fix the $6d$ worldvolume diffeomorphisms by imposing the static gauge
$$
x^\mu=X^\mu, \qquad X^I(x^\mu)\qquad I=6,7,8,9,10
$$
where $X^I(x)$ are five physical scalar fields corresponding to the target--space directions transversal to the M5--brane worldvolume.
We perform the dimensional reduction of three worldvolume directions $x^i$ $(i=3,4,5)$ assuming that the scalar fields $X^I$ and the chiral tensor field $B_{\mu\nu}$  only depend on the three un--compactified coordinates $x^a$ and not on $x^i$.
Then the induced worldvolume metric takes the form
\be\label{reducedm}
g_{\mu\nu}=(\eta_{ab}+\partial_aX^I\partial_bX^I, \,\delta_{ij})\,,\qquad g_{ai}=0\,.
\ee
We use the local gauge symmetry \eqref{33SYM2} to fix the values of the three auxiliary scalars $a^r(x)$ in such a way that the projectors \eqref{PPi} take the form
\be\label{Ppifixed}
P_{\mu}{}^{\nu}=\delta_\mu^a\delta_a^\nu\,,\qquad \Pi_{\mu}{}^{\nu}=\delta_\mu^i\delta_i^\nu\,.
\ee
Then the components $G_{\mu\nu\rho}$ \eqref{GF} of the gauge field strength vanish and $F_{\mu\nu\rho}$ reduce to
\be\label{Freduced}
F_{aij}=\partial_{a}B_{ij}\quad \Rightarrow \quad F_{a}^i=\frac 12 \epsilon^{ijk}F_{ajk}=\partial_a \tilde X^i\,,
\ee
where the dualized components of the gauge field $B_{ij}(x^a)$
$$
\tilde X^i\equiv \frac 12 \epsilon^{ijk}B_{jk}
$$
play the role of the additional three scalar fluctuations of the membrane associated with $D=11$ target--space directions orthogonal to the membrane worldvolume. Indeed, upon the dimensional reduction the M5 brane action \eqref{eq:33actionScaled} becomes
\be\label{M2}
S_{M2}=\int d^3x \sqrt{-\det(\eta_{ab}+\partial_aX^I\partial_bX^I+\partial_a \tilde X^i\partial_b\tilde X^i)}\,,
\ee
which is the action for a membrane in flat $D=11$ space--time in the static gauge.

\section{Conclusion}
Using the non--linear self--duality equation for the 3--form gauge field strength arising in the superembedding description of the M5--brane we have derived a novel form of the kappa--symmetric M5--brane action with a covariant 3+3 splitting of its $6d$ worldvolume.

The value of this action on the mass--shell of the non--linear self--dual gauge field coincides with the on--shell value of the original M5--brane action expressed in terms of the $6d$ scalar function $Q$ of the self--dual chiral field $h_3$ appearing in the superembedding description of the M5--brane. It would be interesting and important to better understand the off--shell relation between the two actions.

Having at hand the M5--brane action in the form \eqref{33M5U}, \eqref{LM5} one can repeat the steps of \cite{Bandos:2008fr} towards understanding the link of this action to the Nambu--Poisson 5--brane of \cite{Ho:2008nn,Ho:2008ve} by restricting the worldvolume pullback of the 11D gauge field $C_3$ to be constant and by partial gauge fixing local symmetries of \eqref{33M5U}, \eqref{LM5} to a group of $3d$ volume preserving diffeomorphisms. The Seiberg--Witten--like map constructed in \cite{Chen:2010br} may be need to relate the fields of the two models. It would be also of interest to relate our construction to a noncommutative M5--brane of \cite{Bergshoeff:2000jn}.

The novel form of the action is also naturally suitable for studying the effective theory of the M5--brane wrapping a $3d$ compact Riemann--manifold.

As another direction of study, one may try, using the superembedding form of the self--duality relation,  to construct an M5--brane action in the form which exhibits 2+4 splitting of the $6d$ worldvolume which may be useful for studying $M5$--branes wrapping $2d$ and $4d$ manifolds, and M5--brane instantons wrapping $4d$ divisors of Calabi--Yau 4--folds in $M_3\times CY_4$ compactifications of M--theory as discussed e.g. in \cite{Witten:1996bn,Kallosh:2005yu,Anguelova:2006sj,Tsimpis:2007sx,Kerstan:2012cy,Bianchi:2012kt}.

\subsection*{Acknowledgements}
The authors are grateful to I. Bandos, F. Bastianelli, M. Cederwall, W.-M. Chen, C.-S. Chu, P.-M. Ho, H. Isono, S. Kuzenko, P. Pasti, I. Samsonov, D. Smith, P. Townsend, M. Tonin, P. West and  B. Zupnik for useful discussions and comments. Work of Sh-L.K. and D.S. was partially supported by the Padova University Research Grant CPDA119349 and by the INFN Special Initiative TV12. D.S. was also supported in part by the
MIUR-PRIN contract 2009-KHZKRX.
Sh-L. K. is grateful to INFN Padova Section and the Department of Physics and Astronomy for kind hospitality and support during his stay in Padova on June 2 - June 13, 2013.
D.S. acknowledges hospitality and support extended to him during the Workshop ``Symmetry and Geometry of Branes in String/M Theory" at Durham University (January 28 - February 1, 2013) at the initial stage of this project, the Galileo Galilei Institute Workshop program ``Higher spins, strings and dualities" (Florence, March 13 - May 10, 2013) and the Benasque Scientific Centre Program ``Gravity - New perspectives from strings and higher dimensions" (July 17-26, 2013) during work in progress.
P.V. is supported by a Durham Doctoral Studentship and by a DPST Scholarship from the
Royal Thai Government.

\appendix

\setcounter{equation}0
\section{Exact check of the M5--brane action non--linear self--duality from superembedding}
To check the form of \eqref{LM5} (or, equivalently, \eqref{LM5fixed}), using the superembedding relations \eqref{F}--\eqref{tG} we should verify that
\bea   \label{eq:SEcheckG}
&\!\!\!\!\!\!\!\!\!\!\!\!\Gt(f,g) = 4Q^{-1}\lrbrk{g - 4g^3 - 4g\tr f^2 + 8\det f} = \frac 1{\sqrt{-\det g_{\mu\nu}}}\frac{\pa \cL_{M5}}{\pa G}\big(F(f,g),G(f,g)\big),&
\eea
and
\bea   \label{eq:SEcheckF}
&\Ft(f,g) = 4Q^{-1}\lrbrk{f(1-4g^2+4\tr f^2) - 8f^3 + 8g f^{-1} \det f}&\nn\\
&= \frac 1{\sqrt{-\det g_{\mu\nu}}}\frac{\pa \cL_{M5}}{\pa F}\big(F(f,g),G(f,g)\big)\,.&
\eea
To verify the above relations, on their right hand sides we should take $G$-- and $F$--derivatives of $\mathcal L_{M5}$ in the form \eqref{LM5fixed}, substitute into the results the expressions \eqref{F} and \eqref{G} for $F$ and $G$ in terms of $f$ and $g$, and to see that they coincide with the left hand sides of \eqref{eq:SEcheckG} and \eqref{eq:SEcheckF}, i.e. with $\tilde G$ and $\tilde F$ expressed in terms of $f$ and $g$.
In particular, we will need to express $\tr(F^2)$, $\tr(F^4)$ and $\det(F)$ in terms of $f$ and $g$.

The algebra is very involved but it is manageable systematically by Mathematica. To this end we used NCAlgebra
package which is found in $\text{http://math.ucsd.edu/$\sim$ncalg/}$.

\subsection{Matrix Notation}
To use Mathematica we should properly define the matrices we deal with.
Let $F_a{}^i$ be the components of the matrix $F$,
$\h_{ab}$ or $\h^{ab}$ be the components of the matrix $\h$ and
$\d_{ij}$ or $\d^{ij}$ be the component of the matrix $\d.$
It will be clear from the context whether the
indices of $\h$ and $\d$ are up or down.
To simplify the notation, we drop $\d$ from all the matrix expressions.

For example, $F_a{}^j\d_{jk}F^k{}_b\h^{bc}F_c{}^i$ is denoted as $F\d F^T\h F$
or just $F F^T\h F.$ This expression is what in previous sections we simply referred
to  as $F^3.$

The inverse matrix $F^{-1}$ has the components $(F^{-1})_i{}^a.$
We will, actually, encounter the adjugate matrix $\adj(F)$
and the cofactor matrix $\co(F)\equiv\adj(F)^T$ more often than
$F^{-1}$ and $(F^{-1})^T.$ The definition of $\adj(F)$
is
\be
\adj(F)_i{}^a \equiv (F^{-1})_i{}^a \det F = \frac 12 \e_{ijk}\e^{abc}F_b{}^j F_c{}^k,
\ee
where
\be
\det F \equiv \ove{6}\e_{ijk}\e^{abc}F_a{}^i F_b{}^j F_c{}^k.
\ee
In the matrix form, the equation \eq{F} reads
\be
F = 4Q^{-1}\lrbrk{f(1+4g^2-4\tr(f^T\h f)) + 8f f^T \h f - 8g\,\h\,\co(f)}
\ee
its transpose is given by
\be
F^T = 4Q^{-1}\lrbrk{f^T(1+4g^2-4\tr(f^T\h f)) + 8f^T \h f f^T - 8g\,\adj(f)\,\h}.
\ee
and
\begin{equation}
Q = 1 - 16 g^4 + 128 g \det(f) - 32 g^2 tr(f^T\eta f) + 16 (tr(f^T\eta f))^2 -
 32 tr(f^T\eta ff^T\eta f).
\end{equation}

We are ready to discuss the computation of the expressions $\tr(F^2)\equiv \tr F^T\eta F$, \linebreak $\tr(F^4)\equiv \tr F^T\h F F^T\h F$ and $\det(F)$ in terms of $f$ and $g$.

\subsection{Outline of computation}
To compute $F^T\h F,$ the following identities are
useful to simplify the results:
\be
\h^2 = 1,
\ee
\be
f\adj(f) = \adj(f)f = \det f,\qquad
f^T\co(f) = \co(f)f^T = \det f,\qquad
\ee
\be
\begin{split}
\adj(f)\h\co(f) &= -(f^T\h f)^{-1}\det(f^T\h f) = -\adj(f^T\h f)\\
                &= -(f^T\h f)^2 + \tr(f^T\h f)f^T\h f-\hlf[(\tr(f^T\h f))^2-\tr((f^T\h f)^2)],
\end{split}
\ee
where in the last equality we used the Cayley--Hamilton
formula for $3\times 3$ matrices.
We also need the Cayley-Hamilton formula of the form
\be
(f^T\h f)^3 = \tr(f^T\h f)(f^T\h f)^2-\hlf[(\tr(f^T\h f))^2-\tr((f^T\h f)^2)]f^T\h f - (\det f)^2.
\ee
Using these formulas one can see that each term in the expression for $F^T \h F$ is proportional
to either
\be
1, \text{\quad or\quad}
f^T\h f, \text{\quad or\quad}
(f^T\h f)^2.
\ee
Therefore,
\be
\begin{split}
\frac{ \tr(F^2) }{16Q^{-2}}=& \text{tr}\left(f^2\right)+
 \left(-48 g \det (f)+16 \text{tr}\left(f^4\right)+8 g^2 \text{tr}\left(f^2\right)-8 \left( \text{tr}f^2\right)^2\right)\\
&+ \left(64 g \det (f) \text{tr}\left(f^2\right)-192 g^3 \det (f)-192 (\det f)^2+96 g^2 \text{tr}\left(f^4\right)+16 g^4 \text{tr}\left(f^2\right)\right.\\
&\qquad\qquad \left.-64 g^2 \left(\text{tr}f^2\right)^2-16 \left( \text{tr}f^2\right)^3+32 \text{tr}\left(f^2\right) \text{tr}\left(f^4\right)\right),
\end{split}
\ee
where $\tr(f^2)$ and $\tr(f^4)$ are shorthand for $\tr(f^T\h f)$ and $\tr((f^T\h f)^2).$

We compute $\tr(F^4)$ and $\tr(F^6)\equiv\tr((F^T\h F)^3)$ using the
same method. We finally trade $\tr(F^6)$ with $\det F$
using the Cayley--Hamilton formula
\be
\det F = \sqrt{-\ove{6} (\tr(F^2)^3 - 3 \tr(F^4) \tr(F^2) + 2 \tr(F^6))}
\ee
The explicit expression for $\det F$ in terms of $f$ and $g$ looks as follows
\begin{eqnarray}\label{detF}
\frac{1}{64}Q^3\det(F) &=& \det(f) + 12 g^2 \det(f) + 48 g^4 \det(f) + 64 g^6 \det(f) +
 192 g \det(f)^2
 \nonumber\\
&& + 1280 g^3 \det(f)^2 - 512 \det(f)^3 -
 4 \det(f) \tr(f^2) - 96 g^2 \det(f) \tr(f^2)\nonumber\\
&&  -  320 g^4 \det(f) \tr(f^2)
 + 256 g \det(f)^2 \tr(f^2) + 4 g (\tr f^2)^2 +
 32 g^3 (\tr f^2)^2\nonumber\\
&& + 64 g^5 (\tr f^2)^2 + 16 \det(f) (\tr f^2)^2 +
 320 g^2 \det(f) (\tr f^2)^2 - 64 \det(f) (\tr f^2)^3\nonumber\\
&& + 64 g (\tr f^2)^4 -
 4 g \tr(f^4) - 32 g^3 \tr(f^4) - 64 g^5 \tr(f^4) - 32 \det(f) \tr(f^4)\nonumber\\
&& -
 640 g^2 \det(f) \tr(f^4) + 128 \det(f) \tr(f^2) \tr(f^4) -
 192 g \,(\tr f^2)^2 \tr(f^4)\nonumber\\
&&+ 128 g (\tr f^4)^2,
\end{eqnarray}

We can now compute the expression in terms of $f$ and $g$ of the term in \eqref{LM5fixed} containing the square root
\bea
&&{\sqrt{1-\frac{    \det (F^2)}{\left(1+G^2  \right)^2}  + \left(G^2+\text{tr}\left(F^2\right)\right)+\ove{2}\frac{  \left(\left(\text{tr}F^2\right)^2-\text{tr}\left(F^4\right)\right)}{1+G^2  }}}  \nn\\
&=& Q^{-3} (1+G^2)^{-1} \sqrt{ \lrbrk{\sum_{n=0}^{12} a_n(f) g^n}^2},
\eea
where the argument of the square root
in the last line, which turns out to form a perfect square, is a polynomial in $g$
with coefficients $a_n(f)$ depending on
$\tr(f^2),\tr(f^4)$ and $ \det(f)$. The form of these coefficients is rather cumbersome, and we do not give it here.
Using the above expressions we can then check that (\ref{eq:SEcheckG}) indeed holds.

We now pass to the check of (\ref{eq:SEcheckF}). In the matrix form it reads
\be
\Ft = 4Q^{-1}\lrbrk{f(1-4g^2+4\tr f^2) - 8f f^T \h f + 8g\,\h\,\co(f)} =\frac 1{\sqrt{-\det g_{\mu\nu}}} \h\frac{\del \cL_{M5}}{\del F^T}\,.
\ee
This is a matrix equation, and we need to compute $F$, $F F^T\h F$, and $\h\,\co(F).$
To do this, we proceed as above and compute
$F$, $F F^T\h F$ and $F F^T\h F F^T\h F$  and then trade
$F F^T\h F F^T\h F$ with $\h\co(F)$ using the relation
\be
\h\co(F)
         = -\frac{\left(F F^T\h F F^T\h F+\frac{1}{2} F \left((\tr F^2)^2-\tr(F^4)\right)-\tr(F^2) F F^T\h F\right)}{\det F}.
\ee
In the final result, the matrices $F$, $F F^T\h F$ and $\h\,\co(F)$
are expressed in terms of $g$, $f$, $f f^T\h f$, and $\h\,\co(f).$ We can then substitute
these into $\del \cL_{M5}/(\sqrt{-det g_{\mu\nu}}\del F)$ which is given by
\be
\frac 1{\sqrt{-det g_{\mu\nu}}}\h\frac{\del \cL_{M5}}{\del F^T}=\frac{   G \h\co(F)}{\left(1+   G^2\right)}
+ \frac{-\frac{ \det(F)\,\h\,\co(F)}{\left(1+  G^2\right)^2}
	   + \frac{   \left( F \text{tr}\left(F^2\right) - FF^T\h F\right)}{1+   G^2} +  F}
{ \sqrt{1-\frac{  \det \,(F^2)}{\left(1+G^2  \right)^2}+  \left(G^2+\text{tr}\left(F^2\right)\right)+\ove{2}\frac{  \left(\left(\text{tr} F^2\right)^2-\text{tr}\left(F^4\right)\right)}{1+G^2 }}}\,,
\ee
\if{}
where we used the basic matrix derivatives
\be
\h\frac{\pa}{\pa F^T}(\tr(F^T\h F)) = \h(2\h F) = 2F,
\ee
\be
\h\frac{\pa}{\pa F^T}(\tr((F^T\,\h F)^2)) = \h(4\h FF^T\,\h F) = 4FF^T\h F,
\ee
\be
\h\frac{\pa}{\pa F^T}\det F = \h co(F).
\ee
\fi
and check that eq. \eqref{eq:SEcheckF} does hold.


\if{}
\bibliographystyle{utphys}
\bibliography{references}
\end{document}
\fi

\providecommand{\href}[2]{#2}\begingroup\raggedright\endgroup

\end{document}